\documentclass[summary]{ursi}

\newcommand\blfootnote[1]{%
  \begingroup
  \renewcommand\thefootnote{}\footnote{#1}%
  \addtocounter{footnote}{-1}%
  \endgroup
}

\usepackage[acronym,shortcuts]{glossaries}
\newcommand\norm[1]{\left\lVert#1\right\rVert}
\newcommand{\rom}[1]{\uppercase\expandafter{\romannumeral #1\relax}}
\usepackage{graphicx}
\usepackage{subcaption} 
\usepackage{amsmath}

\newacronym{b5g}{B5G}{beyond fifth generation}
\newacronym{mmWave}{mmWave}{millimeter wave}
\newacronym{us}{US}{uncorrelated scattering}
\newacronym{wss}{WSS}{wide-sense stationarity}
\newacronym{wssus}{WSSUS}{wide-sense stationarity uncorrelated scattering} 
\newacronym{dpss}{DPSS}{discrete prolate spheroidal sequence}
\newacronym{dsft}{DSFT}{discrete symplectic Fourier transform}
\newacronym{lsf}{LSF}{local scattering function}
\newacronym{v2v}{V2V}{vehicle-to-vehicle}
\newacronym{suv}{SUV}{sport utility vehicle}
\newacronym{los}{LOS}{line-of-sight}
\newacronym{nlos}{NLOS}{non-LOS} 
\newacronym{tf}{TF}{time-frequency} 
\newacronym{lctf}{LCTF}{local channel transfer function}

\newacronym{rms}{RMS}{root mean square}

\title{Stationarity Evaluation of High-mobility sub-6 GHz and mmWave non-WSSUS Channels}

\author{Danilo Radovic*\affref{ref1}, Faruk Pasic\affref{ref1}, Markus Hofer\affref{ref2}, Herbert Groll\affref{ref1}, Christoph F. Mecklenbräuker\affref{ref1}, and Thomas Zemen\affref{ref2} }

\affiliation{%
  \aff{ref1}{TU Wien, Vienna, Austria; e-mail: danilo.radovic@tuwien.ac.at}
  \aff{ref2}{AIT Austrian Institute of Technology GmbH, Vienna}
}

\begin{document}

\maketitle
\blfootnote{The work of D.Radovic and F.Pasic was supported by the Austrian Research Promotion Agency (FFG) via the research project Intelligent Intersection (ICT of the Future, Grant 880830). The work of M. Hofer and T. Zemen was supported by the project DEDICATE (Principal Scientist grant) at the AIT Austrian Institute of Technology.}
\begin{abstract}
Analysis and modeling of wireless communication systems are dependent on the validity of the wide-sense stationarity uncorrelated scattering (WSSUS) assumption. However, in high-mobility scenarios, the WSSUS assumption is approximately fulfilled just over a short time period.  This paper focuses on the stationarity evaluation of high-mobility multi-band channels. We evaluate the stationarity time, the time over which WSSUS is fulfilled approximately. The investigation is performed over real, measured high-mobility channels for two frequency bands, $2.55$ and $25.5$\,GHz. Furthermore, we demonstrate the influence of the user velocity on the stationarity time. We show that the stationarity time decreases with increased relative velocity between the transmitter and the receiver. Furthermore, we show the similarity of the stationarity regions between sub-6\,GHz and mmWave channels. Finally, we demonstrate that the sub-6\,GHz channels are characterized by longer stationarity time. 
\end{abstract}


\section{Introduction}
  High-mobility communication is gaining momentum in the vehicular domain, high-speed railways, uncrewed aerial vehicles (UAVs) communications, etc. However, in order to offer reliable data communication, further understanding of high-mobility wireless channels is required. Furthermore, channel modeling is dependent on the exact characterizations of the actual, measured channels. Since the communication participants are moving at a high velocity, we have to deal with rapidly changing propagation conditions. Therefore, the validity of the \ac{wssus} assumption is limited in both time and frequency. Furthermore, capacity requirements introduce a heavy burden on the crowded sub-6 GHz spectrum. As a possible solution, data offloading in a less characterized \ac{mmWave} spectrum is proposed. Hence, in this paper, we compare the stationarity characteristics of multiband measurements, at $2.55$ and $25.5$\,GHz. It is essential to determine the maximal time duration over which the \ac{wssus} assumption is approximately satisfied. We define this time duration as stationarity time.

  The theoretical approach of statistical characterization of non-\ac{wssus} channels is given in  \cite{1318948, 1532230}. The author introduces \ac{tf} dependent power spectrum representation, known as \ac{lsf}. Furthermore, the author defines the \ac{wssus} as satisfied as long as the \ac{lsf} is constant over a given time and frequency range. Multiple works analyze the stationarity of measured sub-6\,GHz wireless channels, \cite{4475530,6362634, 6553146, 7038157}. They conclude that the minimum stationarity region is on the order of 40\,ms in the time domain. The authors of \cite{7038157} investigate the influence of the communication participants' velocity on the stationarity time. Based on the measurements, they argue that higher relative speed lowers the stationarity time. First stationarity evaluations of high-mobility \ac{mmWave} channels are provided in \cite{Park2019a, Radovic2022}. \cite{Park2019a} analyzes the stationarity time of 28\,GHz channels in highway environment. Their results demonstrate the stationarity time in the range of $2$-$9$\,ms, when the relative speed between TX and RX is $100$\,km/h. The authors of \cite{Radovic2022} show stationarity time of $5$-$16$\,ms for an 60\,GHz urban scenario and the relative speed of $56$\,km/h. Finally, \cite{Tan2017} evaluates multi-band stationarity in a static environment. The authors demonstrate that the channels with a center frequency above 6\,GHz are characterized by shorter spatial  stationarity regions, than conventional channels in the sub-6\,GHz spectrum.

    \textbf{Contributions of the paper:} In this paper we present channel stationarity evaluation of measured high-speed channels. To our best knowledge, this work presents the first stationarity evaluation of real channels, obtained by repeatable measurements. Repeatable measurements mean that we measure the channel multiple times (where we variate just a specific parameter), while moving along the exact same path, and the scattering environment stays constant. Therefore, it enables us to differentiate between the impacts of individual parameters (movement velocity and center frequency).  
  To the best of our knowledge, we present the first evaluation of the influence of the participants’ velocity on the distribution of the \ac{mmWave} spatial stationarity, and the length of the stationarity time. Secondly, we provide a comparison of the channel stationarity for sub-6\,GHz and \ac{mmWave} channels. 
  

\section{Measurement Setup}
\label{Sec:meas_setup}
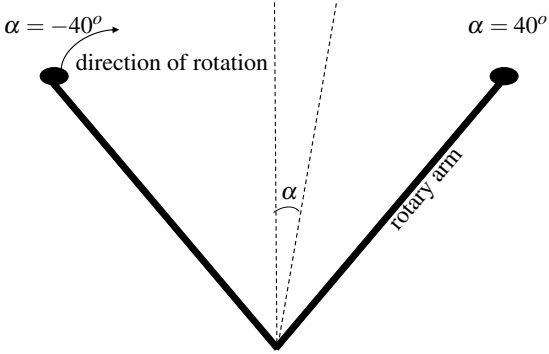
\begin{figure}[htbp]
  \begin{picture}(50,140)
\put(0.5cm,0cm){\includegraphics[width = 0.48\textwidth,page=2,trim=1.5cm 5.8cm 11.9cm 2cm,clip]{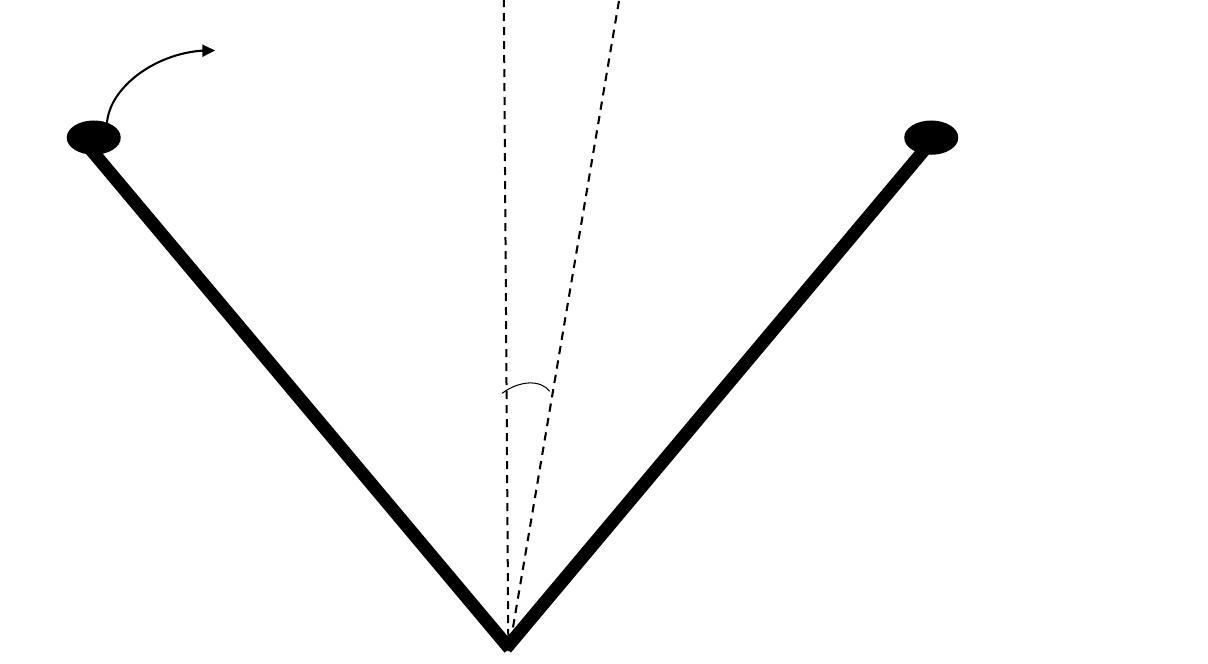}}
\put(0.5cm,4.3cm){\small \text{$\alpha = -40^{o}$}}
\put(6.6cm,4.3cm){\small \text{$\alpha = 40^{o}$}}
\put(4.15cm,2.1cm){\small \text{$\alpha$}}
\put(1.45cm,3.8cm){\small direction of rotation}
\put(5.5cm,1.6cm){\rotatebox{50}{ \small rotary arm }}
  \end{picture} 
  \caption{Measurement setup: TX defined movement.}
  \label{fig:rot_arm}
\end{figure}

In this paper we evaluate measured high-mobility channels. The measurement campaign is performed in a controlled indoor lab scenario. The transmitter is placed on a 1\,m long rotary arm (Fig. \ref{fig:rot_arm}). The measurement is performed once the rotary arm achieves a constant velocity. For each scenario, the channel is measured along the exactly same path, as the rotary arm moves from $\alpha = -40^{o}$ to $\alpha = 40^{o}$ angle relative to the vertical axis. For the multi-band measurements, different transmit antennas are placed at the exact same position. The receiver is fixed in the neighboring room 8\,m away from the transmitter. This setup enables a fair comparison between sub-6\,GHz and \ac{mmWave} systems in terms of small-scale and fast-fading in a high-mobility scenario. In each scenario bandwidth $B=100$\,MHz is set, and contains 500 channel measurements in the time domain. In order to keep a constant number of time samples the subcarrier spacing is set to 400\,kHz and 1\,MHz for the velocity 40\,km/h and 100\,km/h, respectively. In this paper we evaluate the measured channel at $2.55$\,GHz and $25.5$\,GHz and velocity 40\,km/h and 100\,km/h. The measurement setup is explained in detail in \cite{9769375,9900633}.

\section{Stationarity evaluation}
\label{Sec:statEv}
The measured channel, described in Section \ref{Sec:meas_setup},  is defined by its time-variant channel transfer function
\begin{equation} \label{ch_tf}
    {H}[s,q] = H(s T_{\text{s}} , q f_\text{s}),
\end{equation}
where $s\in[1,\cdots,S]$ are time and $q \in [1,\cdots, Q]$ frequency indices, and $T_{\text{s}}$ and $f_\text{s}$ sampling time and frequency. Since the high-mobility channels are non-\ac{wssus}, we define the \ac{tf} subregions, over which we assume stationarity. We denote these regions as \ac{lctf} ${\mathbf{{H}}}_{k_\text{t}} [s',q']$, spanning over $NT_\text{s}$ and $Mf_\text{s}$ length in time and frequency, respectively. $s'$ and $q'$ are local time and frequency indices and $k_\text{t}$ denotes the index of each local region
\begin{equation}
    k_\text{t} \in [1, \cdots, K_\text{t}], \hspace{5pt} K_\text{t} = \left \lfloor \dfrac{S-N}{\Delta_t} \right \rfloor  + 1,
\end{equation}
with time shift $\Delta_\text{t} T_\text{s}$ between two consecutive \ac{lctf}s. When choosing $N$ we have to find a trade-off between the accuracy of stationary evaluation and \ac{lsf} Doppler resolution. By increasing $N$, we risk violating the stationarity assumption, but gain the \ac{lsf} Doppler resolution, which is inversely proportional to $NT_\text{s}$. The channel is measured over a bandwidth of $100$\,MHz, leading to a delay resolution of $10$\,ns. Hence, we assume stationarity over the whole bandwidth and focus on determining the stationarity time. When calculating the \ac{lsf}, we aim to minimize the variance of the spectral estimate. Therefore, we use a multitaper spectral estimator by applying two-dimensional spectral window functions, ${G_\text{w}}[s', q'] = {u_\text{i}}[s'] {\tilde{u}_\text{j}}[q']$, which generates \ac{tf} limited spectral estimates with low sidelobes, as given in \cite{Slepian78}. $\mathbf{u}_i$ and $\mathbf{\tilde{u}}_j$ are \ac{dpss}, described by its energy concentration bandwidth ($2W_\text{t}$, $2W_\text{f}$), and the number of sequences ($I$, $J$), in the time and frequency domain. Here, we deal with a trade-off between the spectral resolution and sidelobe level. We set the $W_\text{t} =2$, $I=2$ and $W_\text{f} =1$, $J=1$ \ac{dpss} parameters in the time and frequency domain, respectively. Further, we define the windowed channel transfer
	\begin{equation}\hat{\mathbf{{H}}}^{(G_\text{w})}_{k_\text{t}} = {\mathbf{H}}_{k_\text{t}} \odot \mathbf{G}_\text{w},
	\end{equation}
	where $\odot$ denotes the Hadamard product. Next, we obtain the windowed Doppler-variant impulse response 
	\begin{equation}
	    \hat{\mathbf{S}}_{k_\text{t}}^{(G_\text{w})} = \mathbf{F}_N \hat{\mathbf{{H}}}^{(G_\text{w})}_{k_\text{t}} \mathbf{F}_M^H,
	\end{equation}
	$\mathbf{F}_i$ and $\mathbf{F}_i^H$ representing discrete Fourier transform (DFT) and inverse DFT (IDFT) matrix of size $i$. We calculate the \ac{lsf} by applying uniform weighting across the $IJ$ windowed Doppler-variant impulse responses
	\begin{equation}
	    \hat{\mathbf{C}}_{k_\text{t}} = \dfrac{1}{IJ} \sum_{w=1}^{IJ}   \left| \hat{\mathbf{S}}_{k_\text{t}}^{(G_\text{w})} \right| ^2 .
	\end{equation}
We define the channel stationarity region as a time period over which the \ac{lsf} is approximately constant. Therefore, we perform a dual-time LSF comparison employing the collinearity spectral distance metric
\begin{equation}
    \mathbf{\gamma}^{(t)}[k_\text{t} , k_{\Delta t}] = \dfrac{ \langle \hat{\mathbf{C}}_{k_\text{t}}, \hat{\mathbf{C}}_{k_{  \text{$\Delta$}   t}} \rangle
    _\mathrm{F} }{ \sqrt{ \norm{ \hat{\mathbf{C}}_{k_\text{t}} }^2_\mathrm{F} \cdot
     \norm{ \hat{\mathbf{C}}_{k_{  \text{$\Delta$}  t}} }^2_\mathrm{F} }},
\end{equation}
where $\langle \mathbf{A}, \mathbf{B} \rangle_\mathrm{F} =\sum_{i,j}{A}[i,j] {B}[i,j]$ and $\lVert\mathbf{A}\rVert_\mathrm{F} = \sqrt{\langle \mathbf{A}, \mathbf{A} \rangle_\mathrm{F}}$ are the Frobenius inner product and Frobenius norm, respectively. $k_\text{t}$ and $k_{\Delta t}$ denote the time indices of the reference and shifted \ac{lsf}, respectively. Collinearity is a bounded metric with values between 0 and 1, where 1 indicates identical and 0 completely dissimilar matrices. For the evaluation we define the channel stationary, as the time duration over which collinearity is above the defined cut-off value 
\begin{align}
    \boldsymbol{t_{\text{stat}}}[k_\text{t}] & = (N+(k_{\Delta t} -1) \Delta_\text{t}) T_\text{s}, \nonumber \\
    & \forall k_{\Delta t}  : \boldsymbol{\gamma^{(t)}}[k_\text{t} , k_{\Delta t}]>0.9.
\end{align}

\section{Results}
\label{sec:results}

In order to establish a comparable movement representation across different velocities, we define  $\alpha \, \widehat{=} \, (k_\text{t}-0.5) \Delta_\text{t} T_\text{s}$ and $\Delta \alpha \, \widehat{=} \, (k_{\Delta t}-0.5) \Delta_\text{t} T_\text{s}$. They denote the angular center position on the arc (Fig. \ref{fig:rot_arm}) of each reference and time-shifted \ac{lsf}. As explained in Section \ref{Sec:statEv}, choosing the size of the \ac{lctf} is a trade-off between the accuracy of the stationarity evaluation and the \ac{lsf} Doppler resolution. 
On one side, choosing a higher value of $N$ can jeopardize the assumption of stationarity over one \ac{lctf}. On the other side, by setting $N$ lower, we decrease the \ac{lsf} Doppler resolution.
With an increased velocity of the TX, Doppler shift increases proportionally. By setting the same angular movement, over which the \ac{lctf} is defined, we achieve the same \ac{lsf} Doppler resolution compared to the maximal Doppler shift, for both velocities. At the frequency band $25.5$\,GHz we set $N=25$ for both velocities, which corresponds to $4^{\circ}$ rotary arm angular movement. For all evaluations presented in this paper, we keep the \ac{lctf} time shift $\Delta_\text{t} = 2$, corresponding to $0.32^{\circ}$ of angular resolution.

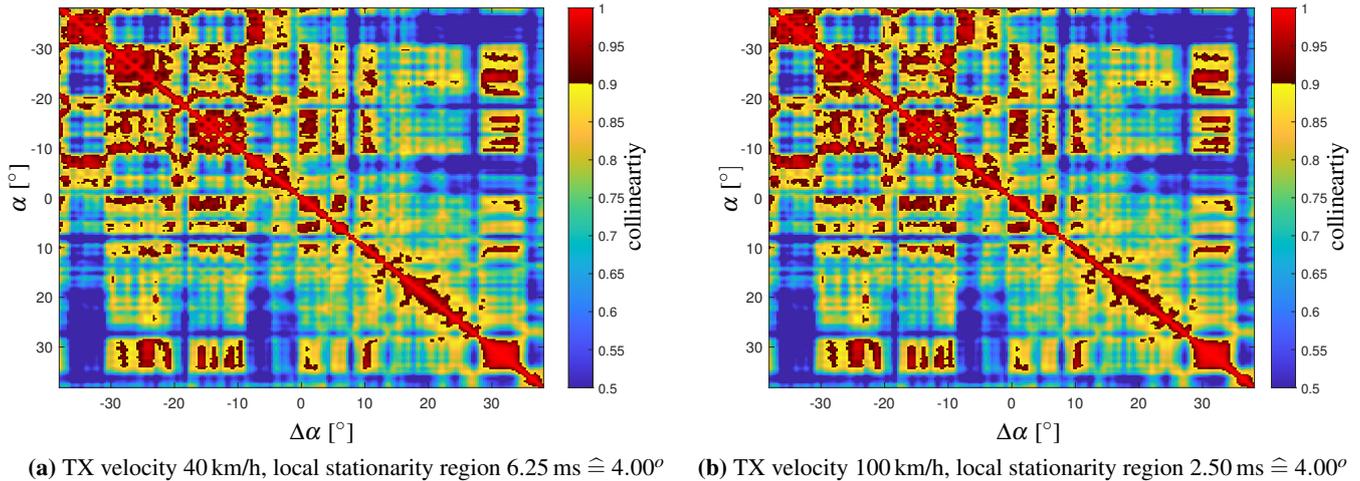
\begin{figure*}[htbp]
\begin{subfigure}{.5\textwidth}
  \begin{picture}(90,179)
		\put(0.3cm,0){\includegraphics[width=0.90\textwidth,trim=0.4cm 0cm 0cm 0cm, clip]{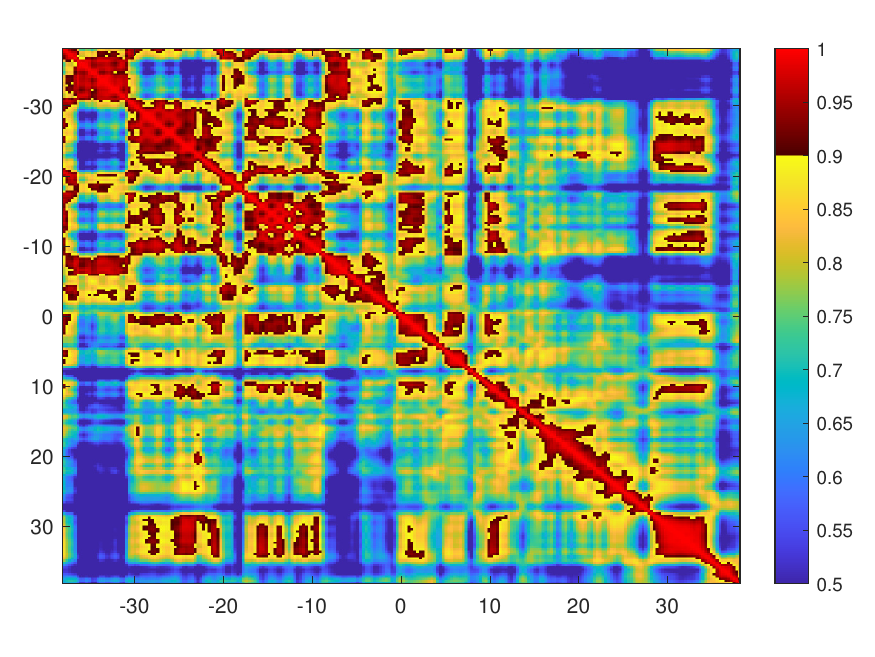}}
	    \put(3.7cm,0cm){\small $\Delta \alpha $ [$^{\circ} $]} 
	    \put(0,3cm){\rotatebox{90}{\small $\alpha $ [$^{\circ} $]}}
	    \put(8.1cm,2.5cm){\rotatebox{90}{\small collineartiy}}
	\end{picture}
		\caption{TX velocity 40\,km/h, local stationarity region 6.25\,ms  $\widehat{=}~ 4.00^{o}$}
		\label{fig:stat25_v40}
\end{subfigure}%
\begin{subfigure}{.5\textwidth}
  
  \begin{picture}(90,179)
		\put(0.6cm,0){\includegraphics[width=0.90\textwidth,trim=0.4cm 0cm 0cm 0cm, clip]{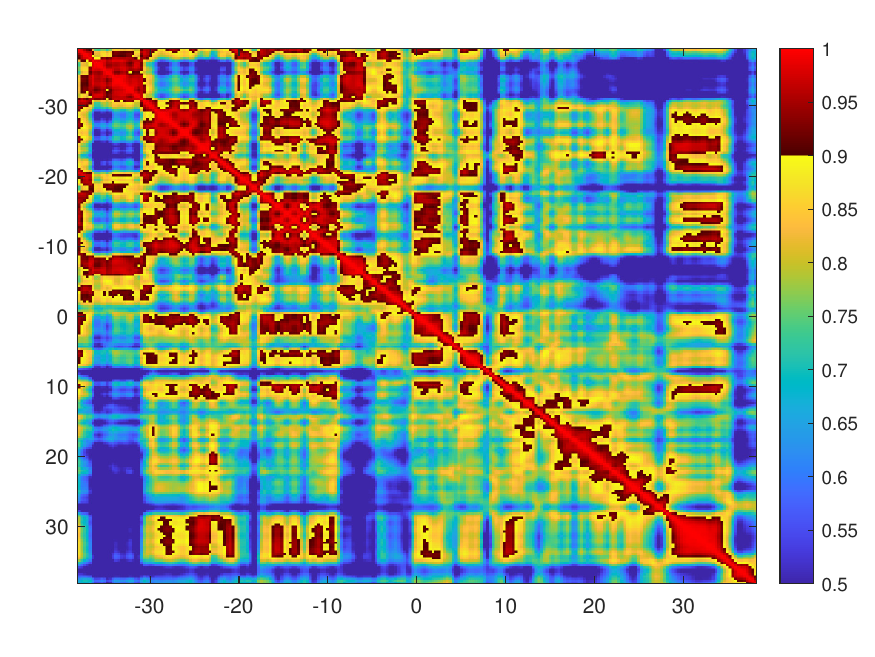}}
	    \put(4.1cm,0cm){\small $\Delta \alpha $ [$^{\circ} $]} 
	    \put(0.45cm,3cm){\rotatebox{90}{\small $\alpha $ [$^{\circ}] $}}
	    \put(8.45cm,2.5cm){\rotatebox{90}{\small collineartiy}}
	\end{picture}
		\caption{TX velocity 100\,km/h, local stationarity region 2.50\,ms  $\widehat{=}~ 4.00^{o}$}
		\label{fig:stat25_v100}
\end{subfigure}
\caption{Stationarity evaluation for f$_\text{c}=25.5$\,GHz,  B=$100$\,MHz.}
\label{fig:wssus_25GHz}
\end{figure*}

\begin{figure}[htbp]
\begin{picture}(90,179)
    \put(-0.1cm,0cm){\includegraphics[width=0.5\textwidth,trim=0.4cm 0cm 0cm 0cm, clip]{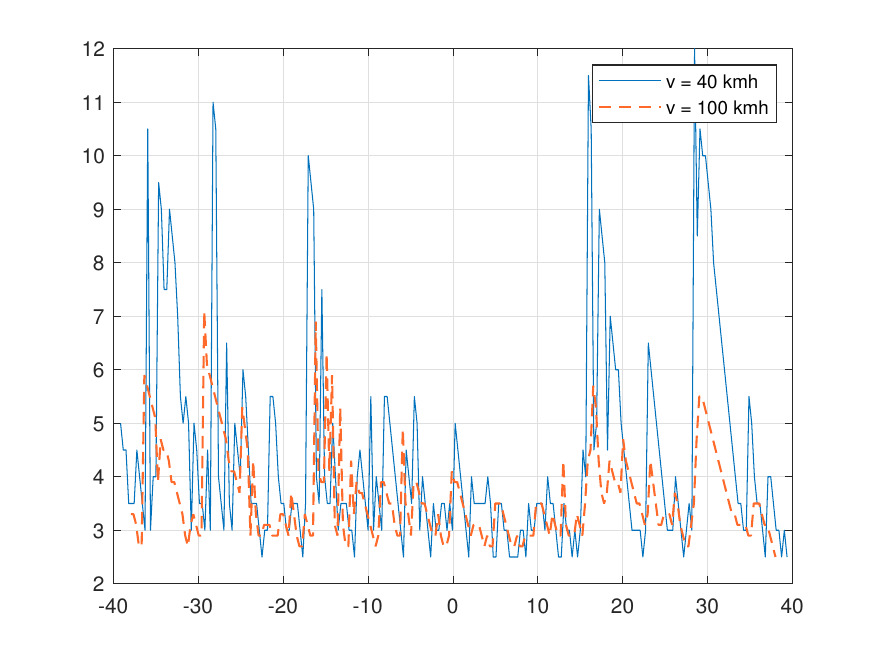}}
    \put(3.7cm,0cm){\small $ \alpha $ [$^{\circ} $]} 
    \put(0cm,2cm){\rotatebox{90}{\small stationarity time  [ms]}}
\end{picture}
\caption{Stationarity time for f$_\text{c}=25.5$\,GHz; TX velocity 40 vs. 100\,km/h; local stationarity region 2.50\,ms.}
\label{fig:wssus_25GHz_40vs100}
\end{figure}

Fig. \ref{fig:wssus_25GHz} shows the stationarity evaluation for $f_\text{c}=25.5$\,GHz. The diagonal shows the collinearity of the \ac{lsf} to itself. As we move along the x-axis, away from the diagonal, we compare the \ac{lsf} with the following \ac{lsf}s in the time domain. Observing the stationarity evaluation for $25.5$\,GHz frequency band for a TX velocity of $40$\,km/h (Fig. \ref{fig:stat25_v40}) and $100$\,km/h (Fig. \ref{fig:stat25_v100}) we notice their high similarity in the spatial stationarity distribution. Fig. \ref{fig:wssus_25GHz_40vs100} shows the comparison of stationarity time between different velocities. For a fair comparison in the time domain, we set the same \ac{lctf} time length of $2.5$\,ms for both velocities. We notice that the stationarity time peaks, visible at lower velocities,
are scaled down, with increased velocity.

Further, we evaluate the stationarity time for the frequency band $2.55$\,GHz. Because the Doppler shift is proportional to the center frequency, sub-6\,GHz experiences a much lower Doppler shift compared to \ac{mmWave}. Therefore, in order to maintain the Doppler resolution high, we increase the length of the \ac{lctf} at the cost of spatial resolution. We set $N=100$, corresponding to $16^{o}$ of angular movement. First, we observe larger stationarity regions compared to the \ac{mmWave} channels. Secondly, we notice similarities between the two frequency bands: 
\vspace{-0.60cm}
\begin{itemize}
    \item low stationarity regions at $\alpha < -27^{\circ}$ and $14^{\circ} < \alpha < 25^{\circ}$, and
    \vspace{-0.3cm}
    \item large stationarity regions $ -27^{\circ} < \alpha < 14^{\circ}$ and $\alpha > 25^{\circ}$.
\end{itemize}
\vspace{-0.5cm}
These similarities between the sub-6\,GHz and \ac{mmWave} band can be explained by the fact, shown in the recent literature, that significant scatterers are visible in both frequency bands \cite{Dupleich2019,Boban2019, Hofer2021}.

\section{Conclusion}
\label{sec:conclusion}
In this paper we evaluated stationarity regions of measured high-mobility multi-band wireless channels. We defined the stationarity region as the maximal time duration, over which the local scattering function (LSF) stays approximately constant. 

The measurement campaign was performed in a controlled lab environment. Such an environment enabled us to perform repeatable measurements and show the influence of the specific parameters. We demonstrated the high similarity of stationarity in the spatial domain when comparing the scenarios where the transmitter was moving at different velocities. Further, that means that the stationarity time is scaled down with an increased velocity. Moreover, we demonstrated the similarities in spatial stationarity between the sub-6\,GHz and millimeter wave (mmWave) wireless channels. Nevertheless, we showed that the sub-6\,GHz channels are characterized by larger stationarity regions in the time domain.

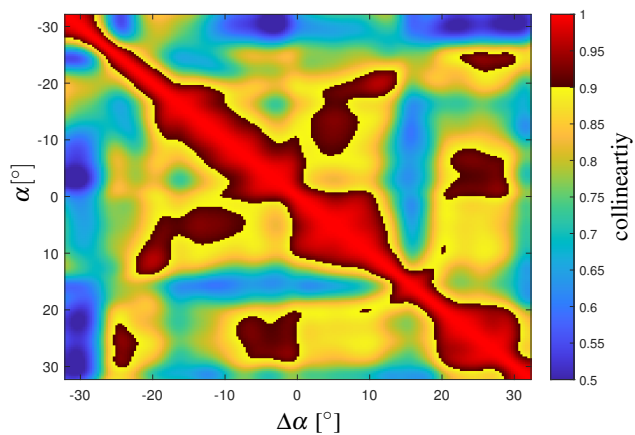
\begin{figure}[htbp]
\begin{picture}(90,170)
    \put(0.5cm,0cm){\includegraphics[width=0.43\textwidth,trim=0.5cm 0cm 0cm 0cm, clip]{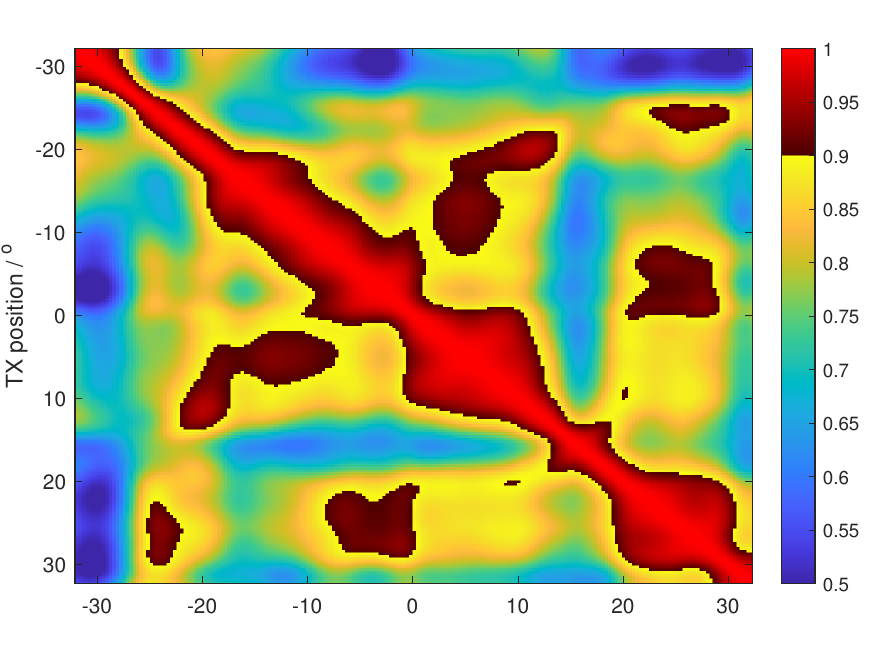}}
    \put(3.7cm,0cm){\small $\Delta \alpha $ [$^{\circ} $]} 
    \put(0.2cm,3cm){\rotatebox{90}{\small $\alpha  [^{\circ}] $}}
    \put(8.1cm,2.5cm){\rotatebox{90}{\small collineartiy}}
\end{picture}
\caption{Stationarity evaluation for f$_\text{c}=2.55$\,GHz; TX velocity 100\,km/h, local stationarity region 10.00\,ms $\widehat{=}~ 16.00^{o}$.}
\label{fig:wssus_2.5GHz}
\end{figure}

\vspace{2.5cm}

\bibliographystyle{IEEEtran}
\bibliography{lib}

\end{document}